\documentstyle[wrapft,seceq,epsf]{ptptex}
\notypesetlogo
\newcommand{\nn}{\nonumber}
\newcommand{\bm}[1]{\mbox{\boldmath $#1$}}

\newcommand{\hs}[1]{\hspace*{#1}}
\newcommand{\tr}{\mathop{\rm tr}}
\newcommand{\diag}{\mathop{\rm diag}}
\newcommand{\wt}[1]{\widetilde{#1}}

\title{
Gauge Symmetry Breaking \\
in Models Inspired by Non-Commutative Geometry}

\author{
Koji {\sc Hashimoto}\footnote{
E-mail: hasshan@gauge.scphys.kyoto-u.ac.jp
}
}
\inst{Department of Physics, Kyoto University, Kyoto 606-8502, Japan} 

\recdate{March 22, 1999}

\abst{
In models inspired by non-commutative geometry, patterns of gauge
symmetry breaking are analyzed, and $SU(5)$ models are found to
naturally favor a vacuum preserving 
$SU(3)_{\rm C} \times SU(2)_{\rm W} \times U(1)_{\rm Y}$. 
A more realistic model is presented, and the possibility of
D-brane interpretation is discussed.
}

\begin{document}
\maketitle


\section{Introduction}

There have been many attempts to study the origin of the Higgs
mechanism. Among these, a method using an idea of non-commutative
geometry (NCG) 
\cite{Connes} was proposed,\cite{non} in which it was shown that the
Weinberg-Salam model can be reconstructed naturally from this
viewpoint. The point of this formulation is to extend the definition
of (gauge) connection so as to be applicable to some non-commutative
space, where the Higgs fields are supplied as a connection of that
discrete space. An intriguing feature of this formulation is that the
Higgs fields naturally develop non-zero vacuum expectation values
(VEVs) and break the gauge symmetry.

The methods of NCG presented by Connes et al.\ are purely
mathematical. Therefore it is of great interest to use their
systematic approach in various systems of physics. G.\ Konisi et al.\
\cite{Saito} proposed a simplified framework for computing Higgs
potential and Yukawa interaction terms derived from gauge kinematics
on a certain disconnected submanifold (this base space is called
$M_4\times Z_N$, where $M_4$ is the Minkowski space-time). Some
phenomenological models have been constructed along this
line.\cite{Saito,SU5saito,LR} \  This framework gives a clear
understanding of the Higgs fields as connection fields, and possesses 
fewer free parameters than NCG. Actually, in usual NCG, the
direction of the gauge symmetry breaking can be chosen freely.

In this paper, utilizing the fact that the method proposed in Ref.\
\citen{Saito} has fewer parameters, we analyze the breaking patterns
of gauge symmetries in some models. After featuring the breaking
patterns of a few models, we show that a simple $SU(5)$ model with an
adjoint Higgs field prefers a vacuum which preserves the standard
model gauge group, $SU(3)_{\rm C} \times SU(2)_{\rm W} \times
U(1)_{\rm Y}$. Then it is
shown to be possible to also obtain the electro-weak symmetry breaking 
under an assumption concerning the coefficients of the Higgs
potential. In the final section we discuss the possibility of
regarding the Minkowski spaces separated from each other as D-branes.


\section{Examples of $M_4\times Z_2$}

According to Ref.\ \citen{Saito}, the potential for the Higgs fields
associated with translation in a discrete space is given as 
\begin{eqnarray}
{\cal V} = {1\over 4}\sum_p \sum_{k,h}\xi_{p,kh} \tr
\left(F_{kh}(x,p)
F^{\dagger kh}(x,p) 
\right) ,
\end{eqnarray} where 
$k$ and $h$ denote the directions of the translation in $Z_N$, and $p$ 
is an element of $Z_N$. The real constant $\xi_{p,kh}$ is a
normalization parameter. The field strengths are defined by use of
Higgs fields $H$ as 
\begin{eqnarray}
F_{kh}(x,p)\equiv H(x,p,p\!+\!k\!+\!h)-
H(x,p,p\!+\!k)H(x,p\!+\!k,p\!+\!k\!+\!h) \ .
\label{def}
\end{eqnarray}

The simplest example is the case with a discrete space $Z_2$. We can
assign gauge symmetries to each element of $Z_2$, and generally they
are $SU(N)$ and $SU(M)$ $(N\geq M)$. Then, only one Higgs field,
$H_i^{\; a}$, in the $({\bf N}, \overline{\bf M})$ representation
appears, and it forms the potential 
\begin{eqnarray}
  {\cal V} = {1\over 4}
\Bigm[\xi_N \tr\!{}_N
\left(F_{kk^{-1}}(x,N) F^{\dagger kk^{-1}}(x,N)
\right)
+\xi_M \tr\!{}_M
\left(F_{k^{-1}k}(x,M) F^{\dagger k^{-1}k}(x,M) \right)
\Bigm] \ ,\nn
\end{eqnarray}
where
\begin{eqnarray}
F_{kk^{-1}}(x,N)\equiv \delta_i^j-H_i^{\; a}H^{\dagger j}_{\;\;\; a}
 \ , 
\quad
F_{k^{-1}k}(x,M)\equiv \delta_a^b-H_i^{\; b}H^{\dagger i}_{\;\;\; a}
 \ . 
\hspace{20mm}
\end{eqnarray}
The index $k$ ($k^{-1}$) represents the translation from $N$ to $M$
($M$ to $N$). When the constants $\xi_N$ and $\xi_M$ are positive, the
vacuum minimizing this potential preserves the gauge symmetry 
$SU(N\!-\!M)\times U(1)^{M-1}$.

In some cases, it is effective to identify some of the vector spaces
$V$ (acted on by the gauge rotation) associated with the elements of
the discrete space, in order to obtain phenomenological models. For an 
example, let us identify the two in the above
\begin{wrapfigure}{r}{6.6cm}
\begin{center}
\leavevmode
\epsfxsize=60mm
\epsfbox{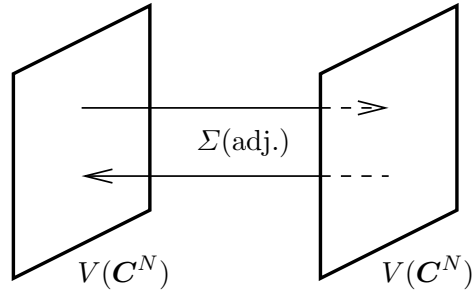}
\put(-100,50){$\Sigma({\rm adj.})$}
\put(-145,0){$V({\bm C}^N)$}
\put(-30,0){$V({\bm C}^N)$}
\caption[]{The adjoint Higgs field is realized as a connection on
  discrete space.}
\label{fig: ncg4}
\end{center}
\end{wrapfigure}
setup (therefore take $N$ equal to
$M$). The Higgs field $\Sigma$ is then in the adjoint representation,
and the Higgs potential is
\begin{eqnarray}
{\cal V}={\xi\over{2}} 
\left( N - 2 \tr\Sigma^2 + \tr\Sigma^4 \right) \ .\hs{10mm}
\label{vacN}
\end{eqnarray}
Referring to the result of Ref.\ \citen{Li}, the vacuum of the
potential (\ref{vacN}) breaks the gauge symmetry $SU(N)$ down to
$SU(N-n)\times SU(n) \times U(1)$, where $n$ is the greatest integer
not larger than $N/2$.


\section{$SU(5)$ model}

A fascinating feature of the $SU(N)$ model (\ref{vacN}) is that in
the case $N\!=\!5$ we obtain the standard model gauge symmetry,
$SU(3)\times SU(2) \times U(1)$. The gauge group $SU(5)$ is one of the
candidates for grand unification, and the adjoint Higgs $\Sigma$ is
just the one needed for the symmetry breaking in the $SU(5)$ GUT.

\begin{figure}[htdp]
\begin{center}
\leavevmode
\epsfxsize=10 cm
\epsfbox{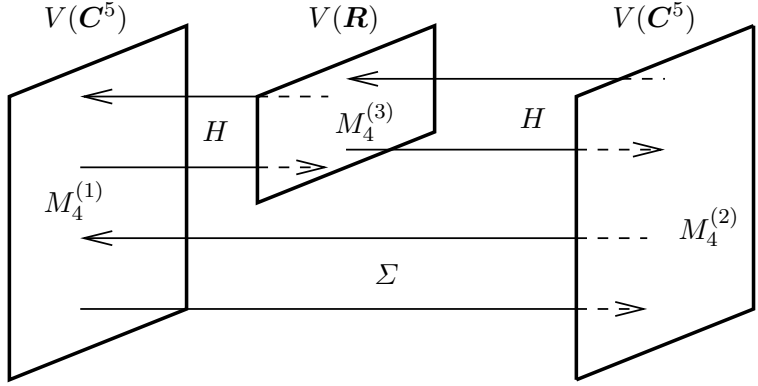}
\put(-210,92){$H$}
\put(-90,97){$H$}
\put(-145,38){$\Sigma$}
\put(-170,135){$V({\bm R})$}
\put(-270,135){$V({\bm C}^5)$}
\put(-55,135){$V({\bm C}^5)$}
\put(-160,95){$M^{(3)}_4$}
\put(-270,65){$M^{(1)}_4$}
\put(-30,55){$M^{(2)}_4$}
\caption[]{A schematic representation for obtaining also the
  five-dimensional Higgs fields.}  
\label{fig: ncg5}
\end{center}
\end{figure}

In the usual minimal $SU(5)$ GUT, one introduces a five-dimensional
Higgs field $H$ in order to realize the electro-weak symmetry
breaking, as well as $\Sigma$. In the context of Ref.\ \citen{Saito},
this matter content is supplied\footnote{Though fermions and their
  interaction terms can be introduced as in Ref.\ \citen{Cham}, we
  concentrate on the Higgs potential in this paper.} by adopting
a space-time $M_4\times Z_3$. This is shown schematically in 
Fig.\ \ref{fig: ncg5}. Vector spaces are assigned as follows: 
$V({\bf C}^5)$ on $M^{(1)}_4$, $V({\bf C}^5)$ on $M^{(2)}_4$ and
$V({\bf R}^1)$ on $M^{(3)}_4$. We have assumed that the vector spaces
on $M^{(1)}_4$ and $M^{(2)}_4$ are the same as before. Hence the
five-dimensional Higgs fields $H$ which connect two $M_4$'s as 
$M_4^{(1)}\!\rightarrow\!M^{(3)}_4$ and
$M_4^{(2)}\!\rightarrow\!M^{(3)}_4$ are the same. \footnote{The gauge 
  fields on $M^{(1)}_4$ and $M^{(2)}_4$ are also the same. The gauge
  field on $M^{(3)}_4$ does not appear, since we put a real vector
  space on it. Thus the gauge group is only $SU(5)$.}

\begin{figure}[tbp]
\begin{center}
\leavevmode
\epsfxsize=120mm
\epsfbox{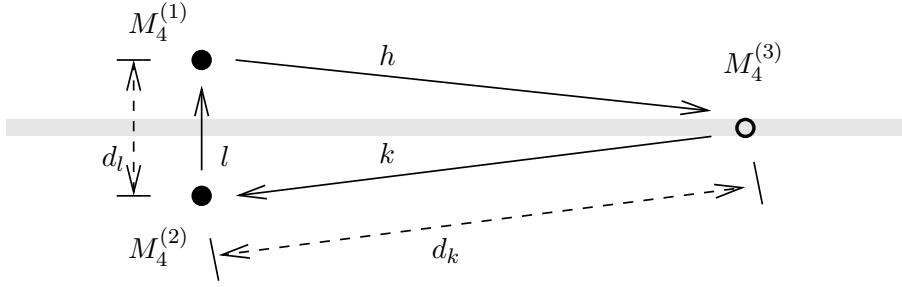}
\put(-200,83){$h$}
\put(-200,46){$k$}
\put(-260,45){$l$}
\put(-180,10){$ d_k$}
\put(-305,45){$d_l$}
\put(-295,95){$M_4^{(1)}$}
\put(-295,10){$M_4^{(2)}$}
\put(-70,80){$M_4^{(3)}$}
\caption[]{The assignment of the distances and the translations
  between different Minkowski spaces. The region above the gray line 
  is identified with the region below.}
\label{fig: ncg6}
\end{center}
\end{figure}

Let us calculate the potential for the Higgs fields. The Higgs field
strengths read from the combinations of the translation in Fig.\ 
\ref{fig: ncg5} are as follows: 
\begin{eqnarray}
\left\{ \begin{array}{l}
 F_{kl}=H^{\dagger a} - H^{\dagger r} \Sigma_r^a,\\
 F_{lh}=H_r - \Sigma_r^a H_a,\\
 F_{hk}=\Sigma_a^r - H_a H^{\dagger r},\\
 F_{l^{-1}k^{-1}}=H_a - \Sigma_a^r H_r, \\
 F_{h^{-1}h^{-1}}=H^{\dagger r} - H^{\dagger a} \Sigma_a^r,\\
 F_{k^{-1}h^{-1}}=\Sigma_r^a - H_r H^{\dagger a},
\end{array}
\right.\hs{10pt}
\left\{ \begin{array}{l}
F_{kk^{-1}}= 1-H^{\dagger r} H_r,\\
F_{k^{-1}k}= \delta_r^s - H_r H^{\dagger s},\\
F_{hh^{-1}}= \delta_a^b - H_a H^{\dagger b},\\
F_{h^{-1}h}= 1-H^{\dagger a} H_a,\\
F_{l \; l^{-1}}= \delta_r^s - \Sigma_r^a \Sigma_a^s,\\
F_{l^{-1}l}= \delta_a^b - \Sigma_a^r \Sigma_r^b.
\end{array}
\right. 
\label{fs}
\end{eqnarray}
Here, the indices $h,k$ and $l$ denote the directions of the
translation (see  
Fig.\ \ref{fig: ncg6}). Note that we have two scale parameters, $d_l$
and $ d_k$, the distance between two Minkowski spaces (the other
parameter $d_h$ is equal to $d_k$ because of the identification of
$M_4^{(1)}$ and $M_4^{(2)}$). 
Since the Higgs fields in the definition (\ref{def}) are
dimensionless, we associate them with the length of the
translation: $\widetilde H(x,p,p\!+\!k)\equiv H(x,p,p\!+\!k)/d_k$. 
Moreover, the Higgs potential should have dimension four. Hence we
make the naive assumption for the normalization parameter that
$\xi_{p,kh} = a_{p,kh} / (d_k^2 d_h^2)$, where $a_{p,kh}$ is a
dimensionless parameter expected to be of order $1$. So as to obtain
the hierarchical structure of the VEVs of the Higgs fields, we assume
$\epsilon\equiv d_l^2 / d_k^2 \ll 1$.

From the relations (\ref{fs}) we obtain the Higgs potential 
as\footnote{The potential (\ref{ssss}) is different from that
  obtained in Ref.\ \citen{SU5saito}. The latter was given through
  another configuration of vector spaces on $M_4\times Z_3$:
\begin{eqnarray}
\lefteqn{{\cal V} ={1\over 4}
\left[
a\left\{\left(\tr(\Sigma^2)\right)^2-\tr(\Sigma^4)\right\}
-(8a-3c)\tr(\Sigma)^2 
\right.
+(b+4c)\left(H^\dagger H\right)^2
-4(b-4d)H^\dagger H 
}\nn\\
&&\hspace{100pt}\left.
+d\left\{H^\dagger\Sigma^2H+ 
\left(\tr(\Sigma^2)\right)\left(H^\dagger H\right)\right\}
+(-6c-12d+e)H^\dagger \Sigma H
\right].\nn
\end{eqnarray}
Though the authors of Ref.\ \citen{SU5saito} introduced five arbitrary
parameters, $a,\cdots, e$, let us associate these with the distances
between the Minkowski spaces, in order to give an
appropriate dimension to the potential:
\begin{eqnarray}
  a=\frac{1}{4 d_l^4}\widetilde a \ , \quad
  b=\frac{2}{4 d_k^4}\widetilde b \ , \quad
  c=\frac{1}{4 d_k^4}\widetilde c \ , \quad
  d=\frac{1}{4 d_l^2 d_k^2}\widetilde d \ , \quad
  e=0 \ .\nn
\end{eqnarray}  
Here $\widetilde a,\cdots,\widetilde d$ are of order $1$. The
parameter $e$ does not stem from the kinetic terms of the Yang-Mills
type, and therefore we set it to zero. The symmetry breaking of the
GUT scale can be seen in the limit $ d_k \rightarrow \infty$. (This 
corresponds to the limit $\epsilon \rightarrow 0$.)
Assuming that the resultant breaking possesses
the appropriate hierarchical structure $\langle\widetilde{H}\rangle
\sim\sqrt{\epsilon}\langle\widetilde{\Sigma} \rangle$ 
($\langle H\rangle \sim\langle\Sigma\rangle$), in this limit we
have only the $a$-term and thus there remains only the adjoint Higgs 
$\Sigma$. 

For positive $a$, the potential of the $a$-part gives rise to a
symmetry-breaking pattern $SU(5)\rightarrow SU(4)\times U(1)$ (see
Ref.\ \citen{Li}), and 
for negative $a$, the potential turns out not to be bounded below. 
Therefore, this model has no
realistic vacuum if the coefficients of the potential are set in the
manner above.
} 
\begin{eqnarray}
\label{ssss}
{\cal V} =\frac{1}{4d_l^4}
\left[{\cal V}_1+\epsilon{\cal V}_2+\epsilon^2{\cal V}_3\right],
\end{eqnarray}
where
\begin{eqnarray}
&&{\cal V}_1=2b_1\left(5-2\tr(\Sigma^2)+\tr(\Sigma^4)\right) \ , 
\quad {\cal V}_2=2(b_2 +b_3)
        (H^\dagger H -2H^\dagger \Sigma H + H^\dagger \Sigma^2H) \ , 
\nn
\\
&&\quad {\cal V}_3=
b_4\left( \tr(\Sigma^2)-2H^\dagger \Sigma H + (H^\dagger H)^2 \right)
+(b_5+b_6)(1-H^\dagger H)^2 + 4b_5 \ .
\label{close}
\end{eqnarray}
Here we have used relations for the coefficients coming
from the identification
\begin{eqnarray}{ll}
&  a_{1,l^{-1}l}=  a_{2,ll^{-1}}\equiv b_1 \ ,\qquad
&  a_{1,l^{-1}k^{-1}}=a_{2,lh}\equiv b_2 \ ,\nn\\
&  a_{3,kl}=a_{3,h^{-1}l^{-1}}\equiv b_3 \ ,\qquad
&  a_{1,hk}=a_{2,k^{-1}h^{-1}}\equiv b_4 \ ,\\
&  a_{1,hh^{-1}}=  a_{2,k^{-1}k}\equiv b_5 \ ,\qquad
&  a_{3,h^{-1}h}=  a_{3,kk^{-1}}\equiv b_6 \ .\nn
\end{eqnarray}
As for the adjoint Higgs, the $(\tr \Sigma^2)^2$ term, which
would be phenomenologically acceptable, does not appear in the 
potential. This is characteristic of our model, due to the structure of 
the field strength (\ref{fs}).

To investigate this potential, it is useful to write the VEV of 
$H$ explicitly as $(h_1,h_2,h_3,$ $h_4,h_5)$ and put
$\langle \Sigma\rangle=\diag(p_1,p_2,p_3,p_4,p_5)$. This is possible
with the use of the $SU(5)$ rotation. When solving the stationary
condition for $\cal V$, the constraint $\tr\Sigma = 0$ should be taken
into account. Thus we add to the above $\cal V$ a term $\lambda \sum
p_i$, where $\lambda$ is a Lagrange multiplier. Then, the necessary
conditions for minimizing the potential $\cal V$ are
\begin{eqnarray}
{{\partial {\cal V}} \over{\partial h_i}}=0
&&\quad\Leftrightarrow \quad
(b_2\!+\!b_3)(1-p_i^2)-\epsilon
\Bigl[
b_4p_i + (b_4\!+\!b_5\!+\!b_6)\sum_j|h_j|^2 +(b_5\!+\!b_6)
\Bigr]=0
\nn\\[-4pt]
&&
\hspace{30pt}
\mbox{or}\quad h_i=0 \ , 
\label{con}\\
{{\partial {\cal V}} \over {\partial p_i}} =0 
&&\quad\Leftrightarrow
\quad
-4b_1(p_i-p_i^3)
-\epsilon\cdot 2(b_2+b_3)|h_i|^2(1-p_i)
\nn\\[-4pt]
&&\hspace{150pt}
+\epsilon^2\cdot b_4(p_i-|h_i|^2)=0 \ ,
\label{cond}\\
{{\partial {\cal V}} \over {\partial \lambda}} =0 
&&\quad  \Leftrightarrow \quad \sum p_i =0 \ .
\label{cond2}
\end{eqnarray}
The pattern of the symmetry breaking of interest is required to
have a hierarchal structure. Thus we add the constraint $\wt{\Sigma} \gg 
\wt{H}$, or in other words,
\begin{eqnarray}
  O(p_i)\gg O(h_i)\sqrt{\epsilon} \ .
\label{hie}
\end{eqnarray}
The vacuum should satisfy Eqs.\ (\ref{con})--(\ref{hie}).

It is natural for the coefficients to be given by 
$b_1\!=\!b_2\!=\!b_3\!=\!b_4\!=\!b_5\!=\!b_6\!=\!1$. 
But adopting these values, it turns out that the potential ${\cal V}$
is minimized at $\langle H\rangle =0$. Hence unfortunately, the
electro-weak breaking does not occur. This is mainly due to the
hierarchical structure of the potential (\ref{ssss}). The global
structure is given by the part ${\cal V}_1$ (leading us to the
GUT breaking), and the next order term ${\cal V}_2$ determines the
breaking pattern for $H$. As seen from the form of the potential
(\ref{close}), ${\cal V}_2$ is bilinear in $H$, and therefore gives
$\langle H\rangle =0$ (or, if the coefficient in front of $H^\dagger
H$ is negative, it gives a large $\langle H\rangle$ together with 
${\cal V}_3$, and the hypothesis of the hierarchy for VEVs is
violated).

Therefore, instead, let us assume the relation $b_2+b_3=0$. Then
setting $b_1\!=\!b_4\!=\!b_5\!=\!b_6\!=\!1$ for simplicity, a
straightforward calculation shows that the whole potential ${\cal V}$
is minimized at
\begin{eqnarray}
&&  \langle \Sigma\rangle=\frac{1}{\sqrt{7}}\diag(-2,-2,-2,3,3) \ ,
\\[-4pt]
&&  \langle H\rangle=\left( 0,0,0,0,h_5 \right) ,\quad 
\mbox{where}\quad |h_5| = \sqrt{\frac23 + \frac{1}{\sqrt{7}}} \ .
\end{eqnarray}
In this way, we obtain a hierarchical structure of the gauge symmetry
breaking of $SU(5)$ GUT: $SU(5)\rightarrow SU(3)_{\rm C} \times 
SU(2)_{\rm W} \times U(1)_{\rm Y} \rightarrow SU(3)_{\rm C}\times
U(1)_{\rm em}$.

\section{Discussion}

In this paper, we have studied the patterns of the gauge symmetry
breaking in models constructed along the proposal of Ref.\
\citen{Saito}. Applying the proposal to a simple $SU(5)$
model\footnote{Minimal $SU(5)$ GUTs in the context of
  NCG\cite{Cham,Song} are completely different from ours.} with an
adjoint Higgs field, the obtained breaking pattern,
$SU(5)\rightarrow SU(3)_{\rm C} \times SU(2)_{\rm W}\times U(1)_{\rm
 Y}$, is found to be very 
preferable phenomenologically. Then, this model was extended
so as to include a five-dimensional Higgs field for electro-weak
symmetry breaking. This is made possible by adding another Minkowski
space, and under the assumption for the coefficients of the potential,
the vacuum breaks the gauge symmetry down to $SU(3)_{\rm C} \times
U(1)_{\rm em}$ with a hierarchical structure.\footnote{We have not
  considered radiative corrections in this paper.} 

Though the connection Higgs formalism seems to be a nice idea to
realize and explain the Higgs mechanism, it suffers from a problem of
non-renormalizability. Actually, this formalism brings about a
coupling reduction, and in general it is impossible to renormalize the
obtained Higgs potential. One way to get rid of this difficulty is to
regard the couplings in the Higgs potential as the one at a certain
energy scale, i.e. the renormalization point.\cite{para} \ But then,
what does this energy scale represent? A possible scenario\cite{ND} is to
think of the Minkowski spaces separated from each other as {\it
  D-branes}, and consider 
the energy scale as the string scale. In string phenomenology,  
new methods for model construction have been exploited with use of 
D-branes, where our four-dimensional space-time is realized as a
worldvolume of the D-brane.\cite{br}

If we regard three parallel Minkowski spaces as D-branes, then the
identification used in this paper and many other models in NCG
becomes very natural. The situation of Fig.\ \ref{fig: ncg6} can be
translated into a configuration of string theory\footnote{A similar
  situation can be found in Ref.\ \citen{braneGUT} in string theory.}
as follows: consider 5 coincident D3-branes (the blobs in 
Fig.\ \ref{fig: ncg6}) and a parallel orbifold singular surface (gray
line) located apart from those D3-branes. The gauge group on the
D-branes is $SU(5)$. (A gauge group $SO(10)$ is obtained when the
location of the orbifold singular surface coincides with that of the
D3-branes.) The 
Higgs field $\Sigma$ can be regarded as a sort of Wilson
line,\cite{Wilson} \ since it is a translational operator. The path
generating a non-trivial Wilson line is the same as the configuration
of a Dirichlet open string in a twisted sector. The scale of this
system is the distance $ d_l/2$ between the D-branes and the orbifold
singularity. Another Minkowski space $M_4^{(3)}$ (the circle) is a
single D-brane on the orbifold singular surface. The scale for
symmetry breaking is determined by the distance between D-branes.

Although on the above points, D-branes provide a natural interpretation
of the configuration of this paper, D-branes are themselves 
supersymmetric objects. In the context of Ref.\ \citen{Saito} and NCG, 
it is difficult to incorporate ${\cal N}=1$
supersymmetry.\cite{chamsusy} \  Therefore, relations between
string theory and NCG should be further clarified (see related
discussion in Ref.\ \citen{Doug}) to understand gauge
symmetry breaking, using recently found non-BPS brane
configurations,\cite{sen} \ for example.


\section*{Acknowledgments}

The author would like to thank S.\ Heusler, T.\ Hirayama,  S.\ Yatsui, 
and K.\ Yoshioka for valuable discussions, and T.\ Kugo for insightful 
comments.  This work is supported in part by a Grant-in-Aid for
Scientific Research from Ministry of Education, Science, Sports and
Culture (\#3160).

\newcommand{\J}[4]{{#1} {\bf #2} (#3), #4}
\newcommand{\andJ}[3]{{\bf #1} (#2) #3}
\newcommand{\AP}{Ann.\ Phys.\ (N.Y.)}
\newcommand{\MPL}{Mod.\ Phys.\ Lett.}
\renewcommand{\NP}{Nucl.\ Phys.}
\renewcommand{\PL}{Phys.\ Lett.}
\renewcommand{\PR}{Phys.\ Rev.}
\renewcommand{\PRL}{Phys.\ Rev.\ Lett.}
\renewcommand{\PTP}{Prog.\ Theor.\ Phys.}
\newcommand{\hep}[1]{{\tt hep-th/{#1}}}
\newcommand{\hepp}[1]{{\tt hep-ph/{#1}}}

\end{document}